\documentclass[conference]{IEEEtran}
\IEEEoverridecommandlockouts

\usepackage{cite}
\usepackage{amsmath,amssymb,amsfonts}
\usepackage{xcolor}
\usepackage{graphicx} 
\usepackage{todonotes}
\usepackage[nolist,nohyperlinks]{acronym}
\usepackage{url}
\usepackage{subfigure}

\usepackage{hyperref}
\hypersetup{
    colorlinks = true,
    allcolors=blue,
    urlcolor=black,
    linkcolor=black
}

\def\BibTeX{{\rm B\kern-.05em{\sc i\kern-.025em b}\kern-.08em
    T\kern-.1667em\lower.7ex\hbox{E}\kern-.125emX}}

\newcommand\citeN\cite

\newcommand\fig[1]{Figure~\ref{fig:#1}}

\newcommand\sect[1]{Section~\ref{sec:#1}}

\newcommand{\FI}[1]{\textcolor{black}{#1}}

\newcommand{\twosubfigeps}[5]{
  \begin{figure}[htb]
    \leavevmode
    \begin{center}
     \subfigure[#2]{
        \label{fig:#1}
        \parbox[t]{0.47\textwidth}{%
            \resizebox{0.45\textwidth}{!}{\includegraphics{figures/#1}}
     \vspace{-1cm}
        }
     }
     \subfigure[#4]{
        \label{fig:#3}
        \parbox[t]{0.47\textwidth}{%
            \resizebox{0.45\textwidth}{!}{\includegraphics{figures/#3}}
     \vspace{-1cm}
        }
     }
    \end{center}
    \vspace{-0.2cm}
    \caption{#5}
    \label{fig:#1_full}
  \end{figure}
}

\newboolean{makevspace}
\newcommand{\cvspace}[1]{%
    \ifthenelse
        {\boolean{makevspace}}
        {\vspace{#1}}
        {}%
    }

\begin{document}

\title{Packet Level Resilience for the User Plane in 5G Networks}
\author{ 
\IEEEauthorblockN{Fabian Ihle\IEEEauthorrefmark{1}, Tobias Meuser\IEEEauthorrefmark{2}, Michael Menth\IEEEauthorrefmark{1}, Björn Scheuermann\IEEEauthorrefmark{2}}
    \IEEEauthorblockA{\IEEEauthorrefmark{1}University of T\"ubingen, Chair of Communication Networks}
    \IEEEauthorblockA{\IEEEauthorrefmark{2}Communication Networks Lab, Technical University of Darmstadt, Germany\\
    Email: \{\textit{fabian.ihle, menth}\}@uni-tuebingen.de, \{\textit{tobias.meuser, bjoern.scheuermann}\}@kom.tu-darmstadt.de}
      \thanks{The authors acknowledge the funding by the Deutsche Forschungsgemeinschaft (DFG) under grant ME2727/3-1,
    by the Federal Ministry of Education and Research of Germany in the project Open6GHub (grant number: 16KISK014), and the LOEWE initiative (Hessen, Germany) within the emergenCITY center.  
The authors alone are responsible for the content of the paper.}
}

\maketitle

\begin{abstract}
The growing demands of ultra-reliable and low-latency communication (URLLC) in 5G networks necessitate enhanced resilience mechanisms to address user plane failures caused by outages, hardware defects, or software bugs.
An important aspect for achieving ultra-reliable communication is the redundant transmission of packets, as also highlighted in 3GPP Release 18.
This paper explores leveraging the Packet Replication, Elimination, and Ordering Function (PREOF) to achieve 1+1 path protection within private 5G environments. 
By extending existing 5G components with mechanisms for packet level redundancy and offloading the reordering mechanism to external servers, the proposed approach ensures minimal packet loss in case of a failure.
A conceptual integration of redundant paths and programmable elements is presented, with considerations for deployment in existing 5G infrastructures and the trade-offs of latency versus enhanced traffic engineering.
Future work aims to evaluate practical implementations using an open source 5G core, P4-based hardware and offloading technologies like DPDK and eBPF.
\end{abstract}


\begin{acronym}
    \acro{FRR}{Fast ReRoute}
    \acro{URLLC}{Ultra-Reliable and Low Latency Communication}
    \acro{PREOF}{Packet Replication, Elimination, and Ordering Function}
    \acro{DetNet}{Deterministic Networking}
    \acro{PTI}{Protection Tunnel Ingress}
    \acro{PTE}{Protection Tunnel Egress}
    \acro{UE}{User Equipment}
    \acro{RAN}{Radio Access Network}
    \acro{GTP-U}{GPRS Tunneling Protocol -- User}
    \acro{gNB}{gNodeB}
    \acro{UPF}{user plane function}
\end{acronym}
\section{Introduction}
In the past years, new applications emerged such as smart factories with industrial machine-to-machine communication, and 5G network slicing with \ac{URLLC} for self-driving vehicles, remote surgery, or drone control.
Those applications require extremely low packet loss and bounded latency~\cite{rfc8578}.
One possible technology to provide connectivity to such applications in a private environment are 5G and beyond networks.
In this paper, we focus on the reliability aspect of \ac{URLLC} in private networks, i.e., reducing packet loss in 5G and beyond communication.

In a 5G network, traffic is forwarded from the \ac{UE}, through the user plane in the \ac{RAN} to the data network.
However, a failure in the user plane is detrimental to connectivity.
Connectivity in the user plane may fail due to power outages, fiber cuts, hardware defects, or software bugs.
Redundant links are therefore a common approach to ensure the connectivity even during failures.
Mechanisms such as \ac{FRR} provide a failover mechanism reacting on a sub-millisecond scale~\cite{MeLi21}.
However, for mission-critical applications, an even smaller restoration time is required.
In 3GPP release 18~\cite{3GPP_Release18}, packet duplication and elimination has been described as possible approach to achieve ultra-high reliability.
Therefore, this work suggests a design for packet duplication and elimination in the context of 5G and beyond that leverages the \acf{PREOF} mechanism introduced in the IETF DetNet working group to protect the user plane of 5G networks with a 1+1 protection scheme.
\ac{PREOF} provides redundant data paths by replicating traffic, sending it over multiple disjoint paths, and eliminating duplicates at the tail end~\cite{rfc9566, rfc8655}.
\section{Background}
In this section, we provide technical background on components in the 5G architecture and the \acf{PREOF}.

\subsection{Components in the 5G Architecture}
Similar to previous generations of cellular networks, 5G-based networks consist of a \ac{UE}, the \ac{gNB} inside the \ac{RAN}, and the core network.
The core network is subdivided into two parts: the control plane and the user plane.
While the control plane is responsible for functions related to network access and management, the user plane forwards packets from authenticated \acp{UE} to the data network, commonly the Internet.
Once a packet from the \ac{UE} is received by the \ac{gNB}, it encapsulates it in a \ac{GTP-U} tunnel and forwards it via a fiber connection to the \ac{UPF}.
The \ac{UPF} is the network function responsible for forwarding user packets to the data network and can be implemented both in hardware and in software~\cite{KMK+22}.

The \ac{UE} is anchored to a specific \ac{UPF}, leading to a disconnection if the \ac{UPF} fails, even if the \ac{gNB} remains functional.
Only after the connection is reset, the \ac{UE} can reconnect to a new \ac{UPF} and continue communication.
\FI{While \acp{UPF} may be redundant, the \ac{UE} must explicitly trigger the reconnection to a different \acs{UPF} on a failure.
Furthermore, the reconnection process to a different \ac{UPF} with an active failover mechanism causes a short disruption in communication leading to a small packet loss~\cite{WeSp24}.}


\subsection{The Packet Replication, Elimination, and Ordering Function (PREOF)}
The \acs{DetNet} working group aims to provide mechanisms that enable real-time applications with extremely low data loss, e.g., in IP/UDP networks~\cite{rfc9566}.
For that purpose, the \acs{DetNet} working group defined a resilience mechanism called \acf{PREOF} which features a 1+1 protection scheme.
The replication function of \acs{PREOF} replicates packets and tunnels them over disjoint links towards the elimination function. 
To that end, the replication function encapsulates packets with sequence numbers and tunnel destination information.
The elimination function eliminates duplicate packets based on the sequence number and forwards them to their destinations.
Furthermore, the ordering function located at the elimination endpoint ensures that forwarded packets are delivered in-order, which requires a packet buffer and adds significant complexity.

A P4-based implementation of the packet replication and elimination functionalities is provided in~\cite{LiMe20} on an Intel Tofino\texttrademark\ with a forwarding speed of 100 Gb/s.
However, the authors emphasize that the objective of their protection mechanism is to protect quickly against path failures and does not compensate for individual packet loss.
Further, the implementation does not provide the ordering function because of limited arithmetic operations and storage access on their hardware target.
To include the compensation for individual packet loss and the ordering function into their approach, a packet buffer is required which is typically not available in P4 switches.
Traffic can be offloaded to an external server that buffers packets and performs the mechanisms.
For offloading, techniques such as eBPF~\cite{HoBr19} or the kernel-bypass Snabb~\cite{PaNi15} can be used.
\section{Concept}
\label{sec:concept}
In this section, we propose a concept of the proposed resilience mechanism leveraging the \acs{PREOF} mechanism in a 5G networking environment.
In \fig{concept_full}, we describe two approaches: providing packet level redundancy for the \ac{UE} with redundant \acp{UPF}, and with redundant \acp{UPF} and \acp{gNB}.

\twosubfigeps{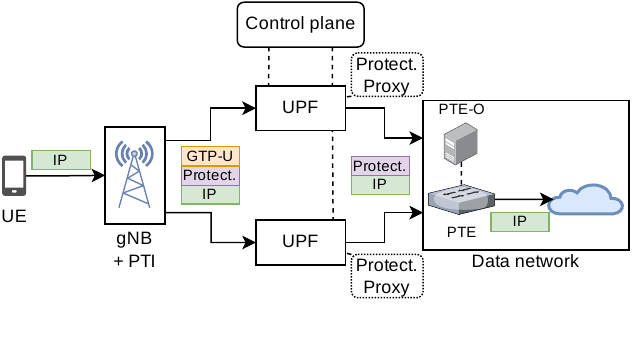}{PTI included in \ac{gNB}.}{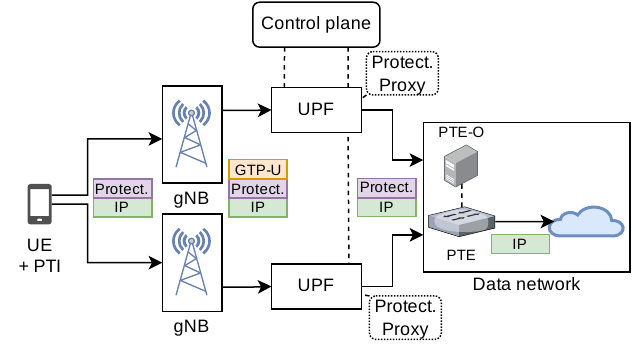}{PTI included in \ac{UE}.}{Overview of a 5G core network incorporating redundant data paths and \acp{UPF} using the \acs{PREOF} mechanism.}

In both approaches in \fig{concept_full}, \acp{UE}, i.e., mobile phones, send their traffic via a wireless link to a \ac{gNB} station in the \ac{RAN}.
A \ac{PTI} node performs the replication and encapsulation mechanism of \acs{PREOF}.
To that end, the \acs{PTI} node adds a protection header to the packet that contains sequence number information and addresses the \ac{PTE} node.
The packets are then replicated by the \acs{PTI} and are sent over multiple disjoint paths towards the \ac{PTE} node.
The \acs{gNB} station encapsulates traffic with a GTP-U header that addresses the corresponding \ac{UPF} according to the 5G standard.
On all disjoint paths, redundant \acp{UPF} with synchronized session states are deployed.
The \acs{UPF} removes the GTP-U header and forwards the packets towards the \acs{PTE} node based on the address information in the protection header.
At the \acs{PTE} node, duplicate packets are eliminated and traffic is forwarded into the data network.
Further, the \acs{PTE} node performs offloading for the ordering function of \acs{PREOF} to an external server, the PTE-O.

The first approach in \fig{concept} incorporates a \ac{PTI} node in the \acs{gNB} station, which performs the replication and encapsulation mechanism of \acs{PREOF}.
Packets are replicated at a single \ac{gNB} station and are sent over multiple disjoint paths and \acp{UPF}.

The second approach in \fig{concept2} includes the \ac{PTI} in the \ac{UE}.
Here, the \ac{UE} performs the protection header encapsulation.
Further, the \ac{UE} sends the encapsulated traffic to multiple \ac{gNB} via multiple wireless links.

For simplicity, only the uplink direction is shown in \fig{concept_full}.
The \ac{PREOF} mechanism is applied in both directions, i.e., from the \ac{gNB} station / the \ac{UE} to the data network and from the data network to the \ac{gNB} station / the \ac{UE}.
In the downlink direction, an edge node of the data network encapsulates packets with the protection header and addresses the PTE, i.e., the \ac{gNB} in \fig{concept}, or the \ac{UE} in \fig{concept2}.
Further, the \acs{UPF} adds the GTP-U header which addresses the \ac{gNB} station.

Typically, the \acs{UPF} terminates the \ac{GTP-U} tunnel and forwards packets based on the underlying Layer 3 information.
However, in large 5G domains, the \acs{UPF} may apply additional traffic engineering mechanisms, such as prioritization with QoS.
This traffic engineering is applied based on the IP header received from the \ac{UE}.
With the proposed protection mechanism, this poses a challenge as the \ac{UE} IP header is not directly accessible by the \ac{UPF} because it is encapsulated by the protection header.
To allow this packet inspection of the \ac{UPF} with \ac{PREOF}, we propose to employ a protection proxy in this case, e.g., using eBPF.
A similar approach using eBPF for an SFC proxy is described in~\cite{HaSt22}.
\section{Discussion}
\label{sec:discussion}
In this section, we discuss the tradeoffs between the two proposed approaches for path redundancy in \sect{concept}.
Further, we discuss the implications of employing a proxy for traffic engineering in the \ac{UPF}, and offloading the ordering function to an external server at the \ac{PTE} node.

In the first approach, including the \acs{PTI} node in the \ac{gNB} protects the communication in case of a \ac{UPF} failure while being transparent to the \ac{UE}.
For this purpose, the \acs{gNB} can either be modified to include the packet replication functionality, or the \acs{PTI} can be a co-located node. 
In this case, the replication is applied after the \acs{gNB} has encapsulated the packet with the GTP-U header.
The former may require changes in the standardization of 5G while the latter does not.
However, including the PTI node in the \ac{gNB} does not protect in case of a \ac{gNB} station failure.
Therefore, in the second approach, the \ac{PTI} is included in the \ac{UE}.
Including the \ac{PTI} in the \ac{UE} does not require changes to the 5G standards as the packet replication can be handled at the \ac{UE} operating system level.
While the second approach protects more points of failure in the 5G network, it also puts more load on the \ac{UE} compared to the first approach.
The \ac{gNB} station typically contains more powerful network components compared to the \ac{UE} for which a packet replication and encapsulation function is more feasible.
Performing such an operation in the \ac{UE} is energy-intensive.
Further, it multiplies the traffic on the wireless link which is not cost-efficient.
Therefore, the second approach, while being more resilient, is more expensive and should only be used for selected applications where \ac{URLLC} is the highest priority.
For both approaches, the \ac{PTE}, i.e., the elimination functionality, is standardized in RFC~9566~\cite{rfc9566}.
The \acs{UPF} does not require any modifications and forwards traffic based on the IP header.

The proposed protection proxy enables traffic engineering mechanisms in the \ac{UPF}.
While the \ac{UPF} could also be modified to ignore \ac{PREOF} headers, thereby eliminating the proxy, such a modification would complicate the deployment of \ac{PREOF} in existing 5G networks.
In contrast, using a proxy increases the latency on the path and adds another point of failure that should be well considered in a \ac{URLLC} environment.
Therefore, the proxy is considered optional in the proposed mechanism.

In addition, we have proposed to offload the ordering function to an external server, as the reordering mechanism is beyond the capabilities of programmable switches such as the Intel Tofino\texttrademark.
However, this also increases the latency and possibly limits the available bandwidth while not being required by every application.
Hence, this mechanism is also considered optional.
\section{Conclusion}
In this work, we proposed to leverage the \ac{PREOF} mechanism introduced in the IETF DetNet working group to achieve packet level redundancy for components in a 5G environment.
For that purpose, existing 5G networking equipment is extended with a packet replication and elimination function.
Further, a packet ordering function is offloaded to an entity in the network to achieve in-order delivery of recombinated packets.
We proposed two approaches: one that includes the packet replication function in the \ac{gNB} station and one that includes it in the \ac{UE}.
To enable traffic engineering capabilities in the user plane, we propose to employ an optional proxy in the \ac{UPF}.
We discussed the tradeoffs of the two proposed approaches, and the implications of the optional proxy and the offloaded ordering function.
In the future, we want to provide an implementation leveraging P4-based hardware switches and offloading technologies such as DPDK and eBPF to further evaluate the implications of the proposed concept.
For the implementation, we consider an open-source 5G core network, e.g., free5GC, coupled with a \ac{RAN} emulator such as UERANSIM or a physical \acs{RAN}.


\bibliography{bibliography/conferences, bibliography/literature}
\bibliographystyle{ieeetr}

\end{document}